\documentstyle[pre,aps,multicol,epsfig]{revtex}

\begin{document}
\draft
\title{Anderson Localization in a String of Microwave Cavities}
\author{C. Dembowski$^{1}$, H.-D. Gr\"af$^{1}$, R. Hofferbert$^{1}$, H. Rehfeld$^{1}$,
        A. Richter$^{1,2}$, and T. Weiland$^{3}$}
\address{$^{1}$Institut f\"ur Kernphysik, Technische Universit\"at Darmstadt,
         D-64289 Darmstadt, Germany\\
         $^{2}$Wissenschaftskolleg zu Berlin, D-14193 Berlin, Germany\\
         $^{3}$Theorie elektromagnetischer Felder, Technische Universit\"at
         Darmstadt, D-64289 Darmstadt, Germany\\
        }
\date{\today}
\maketitle
\begin{abstract}
The field distributions and eigenfrequencies of a microwave resonator which is
composed of 20 identical cells have been measured. With external screws the
periodicity of the cavity can be perturbed arbitrarily. If the perturbation is
increased a transition from extended to localized field distributions is
observed. For very large perturbations the field distributions show signatures
of Anderson localization, while for smaller perturbations the field
distribution is extended or weakly localized. The localization length of a
strongly localized field distribution can be varied by adjusting the
penetration depth of the screws. Shifts in the frequency spectrum of the
resonator provide further evidence for Anderson localization.
\end{abstract}


\begin{multicols}{2}
\narrowtext

\section{INTRODUCTION}
\label{Introduction}

In 1958 P.W. Anderson calculated the effects of perturbations of a periodic
lattice on the eigenvalues and eigenfunctions of the Schr\"odinger equation
\cite{Anderson}. He was able to show that for a variety of disordered
potentials the eigenfunctions are exponentially localized in a small region of
the lattice due to the interference of waves scattered from the perturbations
or impurities. Till then the localization of waves played a crucial role in
almost every physical domain -- for a review see e.g. \cite{Tiggelen}. In the
case of a long, disordered one-dimensional (1D) chain it could be shown
rigorously, that Anderson localization occurs for a very large class of
potentials \cite{Pastur,Kunz,Delyon}. However there are classes of disordered
potentials where extended states do exist \cite{Denbigh,Crisanti}. It has been
conjectured that localization effects are far more general
\cite{John,Azbel,Kirkpatrick} and are a generic feature of wave equations
\cite{Figotin1,Figotin2}.

Experiments working with electromagnetic waves are usually focusing on
secondary features of localization, e.g. coherent backscattering
\cite{Albada,Wolf} or the investigation of the transmission and absorption
coefficients \cite{Garcia,Genack,Stoeck3}. A direct search for localization in
electromagnetic field distributions has only very recently been attempted in
two-dimensional (2D) \cite{McCall} and three-dimensional (3D) systems composed
of 1D waveguides \cite{Zhang}. In the present paper we like to report on the
direct observation of Anderson localization in an electromagnetic system, i.e.
a microwave resonator, governed by the {\it vectorial} Helmholtz equation, with
a perturbed periodicity. So far microwave resonators have been a major
experimental tool for the investigation of so-called quantum billiard systems
\cite{Stoeck1,Sridhar,PhysRevLet,PhysRevE,PhysRevLet2,PhysRevE2,PhysRevE3},
i.e. a point-like particle caught in a potential with infinitely high walls, or
recently for the study of models used in nuclear physics \cite{PhysRevLet4}. In
all these experiments the analogy between the {\it scalar} Helmholtz equation
which describes the electromagnetic field inside a {\it flat} microwave cavity
and the Schr\"odinger equation is used and the energy spectra of the resonators
are statistically analyzed -- for a review of a wide range of experiments and
the statistical methods used see e.g. \cite{RichterBuch}.

Here we will discuss experiments performed with a so-called three-dimensional
microwave cavity, i.e. a resonator that has to be described by the vectorial
Helmholtz equation and belongs to the class of resonators investigated in
\cite{Deus,PhysRevE4,PhysRevLet3,Stoeck2}. As mentioned these experiments
compose to our knowledge the first and direct experimental search for Anderson
localization in such systems, not to be confused with the study of higher
dimensional lattices still described by the Schr\"odinger equation
\cite{Pichard}. One of the main problems in such an experimental investigation
is the limited size of the system at hand. In numerical simulations of finite
chains described by the Schr\"odinger equation the number of elementary cells is
usually some hundred cells or even higher (for an early review see e.g.
\cite{Ishii}), while a realistic experimental set-up with a chain of microwave
resonators has to have a much smaller number of elementary cells to keep the
dimension of the system at an acceptable level.

The paper is organized as follows. In Sect.~\ref{experiment} we will give a
description of the microwave resonator and the experimental methods used to
measure the resonance frequencies and the field distributions of the cavity.
After having verified that the unperturbed system is indeed periodic and that
we find extended states we proceeded to investigate the influence of a single
perturbation of the periodicity on the eigenstates and eigenfrequencies -- the
results are given in Sect.~\ref{singledist}. As a third possible configuration
we tried to set-up a disordered chain and measured the field distributions
discussed in Sect.~\ref{multiple}.

\section{EXPERIMENT}
\label{experiment}

\begin{figure}[bht]
\centerline{\epsfxsize=8.6cm \epsfbox{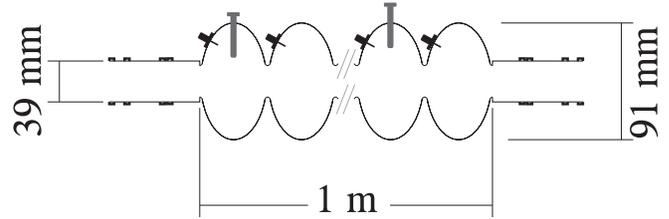}} \vspace*{2.5ex}
\caption{Sketch of the two end sections of the modified accelerating cavity.
Besides the identical dipol antennas two of the adjustable lead screws are
shown.} \label{cavity}
\end{figure}

To simulate periodic and perturbed systems, an accelerating cavity of the
superconducting Darmstadt linear electron accelerator S-DALINAC
\cite{S-DALINAC} has been modified. The cavity itself is manufactured from 2~mm
thick Niobium sheetmetal and consists of 18 identical cells and two slightly
different cells at the ends of the chain to compensate the influence of the
cut-off tubes attached to the outer cells. The cylindrical symmetric section
with the 20 cells has a length of 1m and the diameter of a single cell varies
between 91~mm and 39~mm. A sketch of the modified accelerating cavity is given
in Fig.~\ref{cavity}. Below 3.5~GHz the two cut-off tubes cause an exponential
decay of the electromagnetic field outside the 20 cell section \cite{Jackson}
and above 3.5~GHz two Niobium plates can be used to close the first and the
last cell. In both cases only the 20 cell section is excited up to 20GHz by a
HP8510B vectorial network analyzer connected to a set of 20 identical and
periodically mounted capacitively coupling dipole antennas. We chose a large
penetration depth of 8~mm for the antennas to ensure that even modes which are
localized in just one or two cells can be observed in the transmission spectra.
Care has been taken to ensure that all antennas are identical and periodically
mounted so that the periodicity of the resonator is not perturbed by the
antennas themselves. To perturb the periodic setup every second cell is
equipped with an adjustable lead screw with a diameter of 20~mm which can
penetrate into the resonator's volume. The penetration depth of each screw
$d^i=0{\rm mm},\ i=1,3,\ldots,19$ can be continuously varied between 0~mm and
47~mm. The cavity therefore allows the investigation of three different
set-ups, i.e. a periodic system, a system with a single perturbation or
impurity with variable strength and a disordered system where the $d^i$'s are
set to random values. Despite the fact that the measurement of the field
distributions has been performed at room temperature, all materials used will
become superconducting at temperatures which can easily be reached inside a
LHe-bath cryostat described in \cite{RichterBuch} to allow future high
resolution measurements of the cavity's spectra.

The electromagnetic fields $\vec{E}$ and $\vec{B}$ inside the cavity are
described by the vectorial Helmholtz equation \cite{Jackson}
\begin{equation}
(\Delta + \epsilon \mu \frac{\omega^2}{c_0^2})\vec{E}(\vec{r}) = \vec{0}
\end{equation}
and
\begin{equation}
(\Delta + \epsilon \mu \frac{\omega^2}{c_0^2})\vec{B}(\vec{r}) = \vec{0}
\end{equation}
with the corresponding boundary conditions
\begin{equation}
\vec{E}_\parallel (\vec{r}) \big|_{\partial G} = \vec{0} {\rm \qquad and
\qquad} \vec{B}_\perp (\vec{r}) \big|_{\partial G} = \vec{0}
\end{equation}
on the walls $\delta G$ which are assumed to be ideally conducting. The well
known analogy between flat electromagnetic cavities and two-dimensional quantum
potentials, that has stimulated a multitude of experiments \cite{RichterBuch}
is of course lost. As mentioned above the effects of Anderson localization can
be found either in the eigenfunctions, in our case the electromagnetic field
distributions, or the energy spectrum, in our case the set of resonance
frequencies, of a system. We examined therefore both, the field distributions
and the energy spectrum, for signatures of localization effects.

The electromagnetic field amplitudes inside the cavity can be measured using
Slater's theorem formulated in \cite{Slater} that describes shifts of the
resonance frequencies of a microwave cavity if a perturbing body is brought
into the resonating volume. The electric and the magnetic field energies inside
a microwave cavity which is excited in resonance are of equal magnitude if the
cavity is in a stationary state. If an external body is introduced into the
resonator, the field energies are shifted and the resonance frequency $f_0$
will be adjusted in a way that the electric and magnetic field energies become
equal again and a new stationary state is reached. According to \cite{Slater}
the frequency shift $\Delta f = f - f_0$ of a resonance caused by a perturbing
body of volume $\Delta V$ is
\begin{equation}
\frac{\Delta f}{f_0} = \frac{1}{4U}\int\nolimits_{\Delta V}(\epsilon\epsilon_0
\vec{E}^2 - \mu\mu_0 \vec{H}^2)dV, \label{Slatereq}
\end{equation}
where $U$ is the total energy stored in the electromagnetic field and
$\epsilon$ and $\mu$ are the perimittivity and the permeability of the
perturbing body. If $\Delta V$ is small compared to the wavelength of the mode
under investigation, the perturbing body is moved on a nodal line of the
magnetic field (i.e. $\vec{H}=\vec{0}$) and is composed of a dielectric
material, Eq.~(\ref{Slatereq}) can be written as
\begin{equation}
\frac{\Delta f}{f_0} = \frac{\epsilon \epsilon_0 \Delta V}{4 U} \vec{E}^2,
\end{equation}
which immediately leads to the proportional relation
\begin{equation} \label{FpropE}
\Delta f \propto \vec{E}^2.
\end{equation}
The squared field strength $\vec{E}^2$ can therefore be calculated directly
from measurements of the frequency shift caused by a small perturbing body.

\begin{figure}
\centerline{\epsfxsize=8.6cm \epsfbox{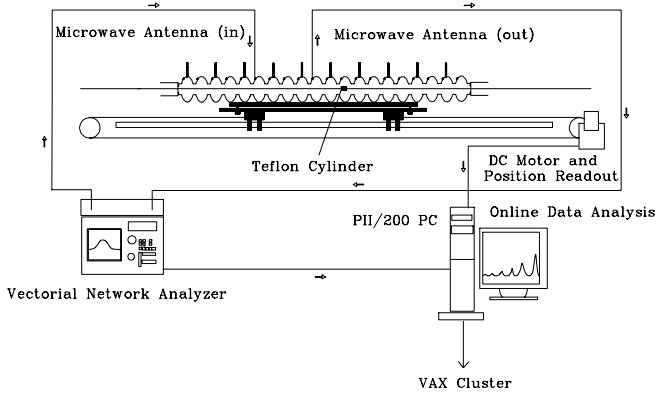}} \vspace*{2.5ex}
\caption{Sketch of the experimental setup we used to measure the field
distributions of the resonator. The cavity is moved by a DC motor on a sledge
while the Teflon cylinder is fixed on the axis. A Pentium-133 PC is used for an
online data analysis.} \label{setup}
\end{figure}

To measure the frequency shift we used the experimental set-up sketched in
Fig.~\ref{setup}. The resonator is moved with a speed of $1 {\rm m/min}$ while
the perturbing body is fixed on the symmetry axis of the cavity with a very
thin string, to prevent oscillations of the body. The first twenty resonances
of the cavity are transverse magnetic (TM) modes for which the magnetic field
is zero on the axis. We further used a cylindrical Teflon bead as a perturbing
body with a volume of $\Delta V = 2\rm{ mm}^3$ and a permittivity of
$\epsilon\approx 2.1$ and a permeability of $\mu\approx 1$ \cite{Teflon}. The
special symmetry of our system and the perturbing body we are using allows us
to measure $\vec{E}^2$ directly by using Eq.~(\ref{FpropE}) unlike similar
experiments \cite{Stoeck2}, where a metallic perturbing body is used and the
composed quantity $-2 \vec{E} + \vec{B}$ is measured.

\begin{figure}[hbt]
\centerline{\epsfxsize=8.6cm \epsfbox{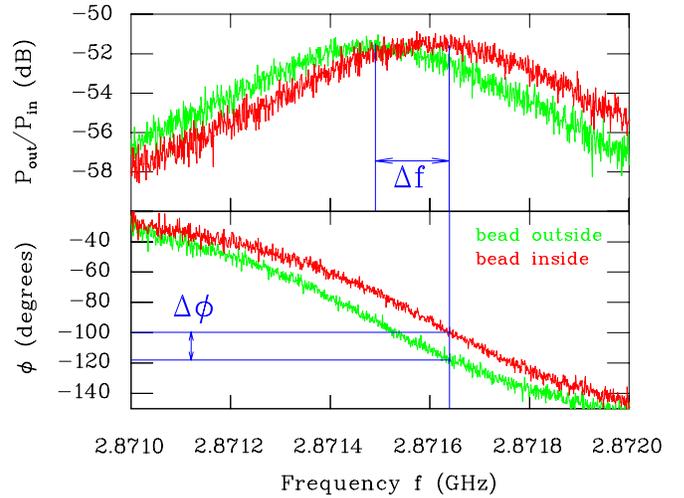}} \vspace*{2.5ex}
\caption{Transmission spectra (top) and phase relations $\phi$ between the
received and the emitted wave (bottom), showing the perturbation introduced by
the Teflon cylinder.} \label{phaseshift}
\end{figure}

The network analyzer can be set to continuously sweep a range of 1~MHz across
the last position of the resonance which takes a time of 184ms. In the upper
part of Fig.~\ref{phaseshift} a part of the transmission spectrum with a
resonance at 2.8715~GHz and the Teflon cylinder outside and inside the
resonator is shown. The frequency shift $\Delta f$ caused by the perturbation
of the electromagnetic field is clearly visible. Nevertheless, additional noise
blurs the position of the resonance so that it is impossible to find its
position by simply looking for the frequency with the highest transmission.
Transmitting the data  via the IEEE-bus for an offline analysis with e.g.
fitted resonance curves is also not practical since during the time needed for
the data transfer ($\approx 1\rm s$) the cavity moves about $16{\rm mm}$.
Beside the measurement of the ratio between the received and the emitted wave,
the HP8510B allows the analysis of the phase relation $\phi$ between the
received and the emitted wave which shifts by $\pi$ at the resonance frequency.
For a resonance with a frequency $f_0$ and a quality factor $Q$ one can write
the phase relation as \cite{Jackson}
\begin{equation}
\phi=\arctan(\frac {f f_0} {Q (f_0^2 - f^2)}) + \pi/2 \label{phaseshiftexact}
\end{equation}
which can be expanded around the point $f = f_0$ to a linear relation
\begin{equation}
\phi=\frac{\pi}{2} + 2 \frac {Q(f_0-f)}{f_0} + {\it 0}(f^2)
\label{phaseshifteq}
\end{equation}

As one can see in the lower part of Fig.~\ref{phaseshift} the frequency shift
$\Delta f$ is small compared to the region where the linear series
approximation of Eq.~(\ref{phaseshiftexact}) is applicable. The network
analyzer is therefore set to emit a wave with the frequency $f_0$ and measure
the phase relation $\phi$ between the received and the emitted wave. Since no
time consuming sweep is required for measurements performed at a constant
frequency, the Teflon bead is assumed to be fixed during each measurement and
the received signal can be averaged up to 4000 times, therefore greatly
reducing the underlying noise. The difference of the phase relations with the
bead outside and inside the cavity, $\Delta \phi,$ as shown in the lower part
of Fig.~\ref{phaseshift}, can then easily be used with Eq.~(\ref{phaseshifteq})
to compute the field amplitudes. From Eqs.~(\ref{FpropE}) and
(\ref{phaseshifteq}) one gets the proportionality relation
\begin{equation}
\Delta \phi \propto \Delta f \propto \vec{E}^2. \label{phaseshiftprop}
\end{equation}

\begin{figure}[hbt]
\centerline{\epsfxsize=8.6cm \epsfbox{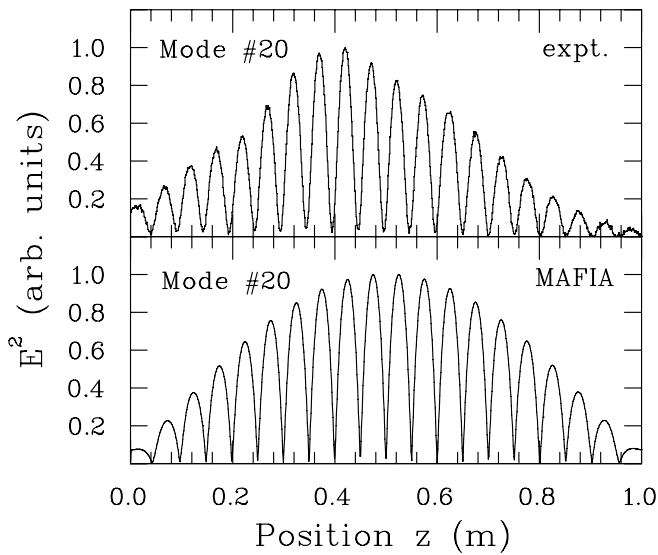}} \vspace*{2.5ex}
\caption{Field distributions $\vec{E}^2(z)$ of the unperturbed cavity, measured
with our setup (expt.) and calculated with the finite element code MAFIA. The
position on the axis of the cavity is denoted by $z$.} \label{undisturbed}
\end{figure}

In a first step we measured the field distribution of the unperturbed cavity,
i.e. the penetration depths of the screws were set to zero, and compared our
measurements with finite element calculations using the MAFIA \cite{MAFIA}
computer code. The model used for the finite element calculations is
cylindrically symmetric with a resolution of $1{\rm mm}\times 0.25{\rm mm}$ and
allows the calculation of the resonance frequencies and the field distributions
of the first 40 TM modes. In Fig.~\ref{undisturbed} the experimental and the
calculated electric field distributions for the 20th mode are compared, showing
that the effects of small deviations from the ideal geometry, either due to the
additional holes that were drilled into the cavity or the cut-off tubes, are
negligible.

\section{A SINGLE PERTURBATION}
\label{singledist}

\begin{figure}[hbt]
\centerline{\epsfxsize=8.6cm \epsfbox{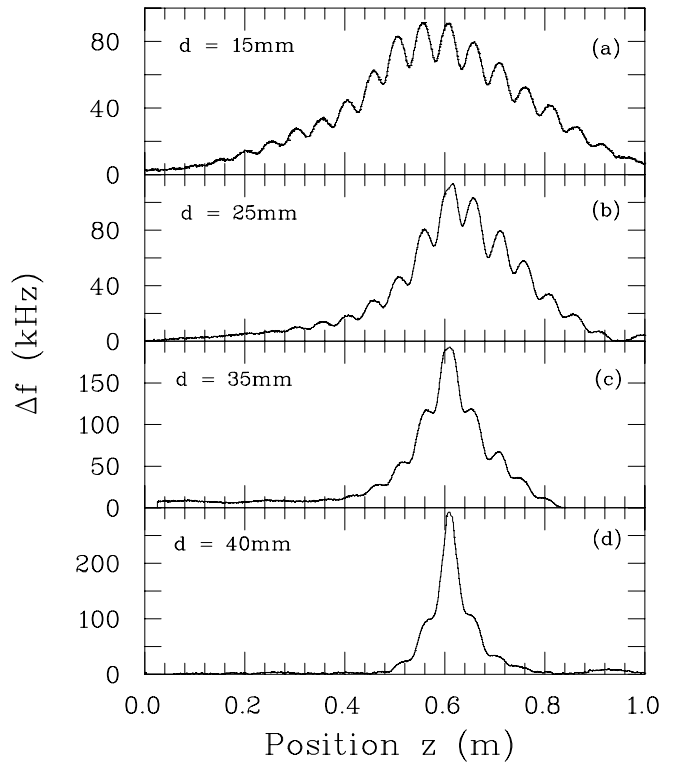}} \vspace*{2.5ex}
\caption{Shift $\Delta f$ in the resonance frequency of the first resonance
caused by the Teflon cylinder for different penetration depths $d^{13}$ of the
screw in cell \#13.} \label{single}
\end{figure}

If the system does show a transition from extended to localized states if the
periodicity is perturbed -- despite the finite size and the fact that it has to
be described by the vectorial Helmholtz equation -- one expects for a single
perturbation of the otherwise periodic chain only {\it one} local impurity mode
in each band with a localized field distribution. We therefore investigated the
development of the field distribution of the first resonance if the penetration
depth of an arbitrary screw (here the one in cell \#~13) is increased. The
evolution of the wave function is shown in Fig.~\ref{single}, where we plotted
the frequency shift $\Delta f$ against the position $z$ of the Teflon bead.
Beside perturbing the periodicity of the cavity the screw also breaks its
cylindrical symmetry so that the cavities axis is not a nodal line of the
magnetic field anymore. Calculations with MAFIA have nevertheless shown that
the magnetic field on the axis is still zero for a wide range of penetration
depths. For penetration depths above approximately $40{\rm mm}$ the actual
electric field is slightly larger than the one determined by
Eq.~(\ref{phaseshiftprop}) since the proportionality relation has to be written
as
\begin{equation}
\Delta \phi \propto \Delta f \propto \epsilon\epsilon_0\vec{E}^2 -
\mu\mu_0\vec{H}^2.
\end{equation}
For $d^{13}=15{\rm mm}$ the first resonance excites the whole cavity
(Fig.~\ref{single}a) while for an increasing $d^{13}$ the field amplitude drops
down to zero almost everywhere in the resonator. The field distribution of the
first resonance even localizes up to a point where {\it only} the disturbed and
a few neighboring cells are excited (Fig.~\ref{single}d). As it is the case for
systems described by the Schr\"odinger equation, the field distributions of the
other 19 modes of the first TM band are still extended over the whole cavity.

In the case of strong localization in quantum systems the envelope of the
localized eigenfunction can for a wide range of perturbed potentials be
approximated by an exponentially decaying function \cite{Anderson,Pastur}, i.e.
\begin{equation}
|\psi_{loc} (z)|^2 \propto \exp \left( - \frac{|z - z_0|}{L_{loc}} \right),
\label{loclengtheq}
\end{equation}
where $z_0$ is the position of the perturbation and $L_{loc}$ is called the
localization length.

\begin{figure}[hbt]
\centerline{\epsfxsize=8.6cm \epsfbox{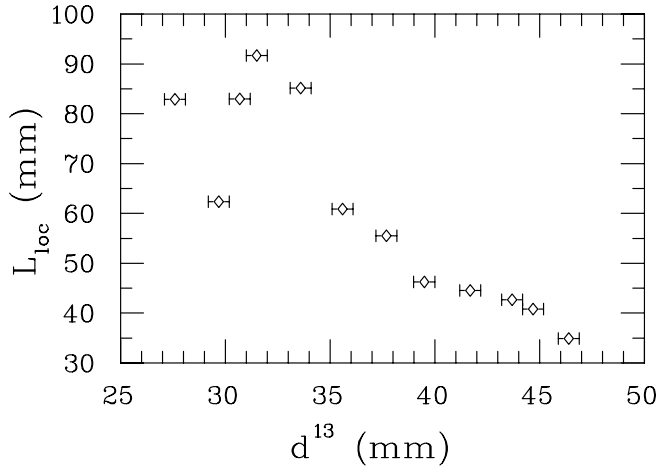}} \vspace*{2.5ex}
\caption{Localization length $L_{loc},$ i.e. the length where $\vec{E}^2$
decreases by $1/e,$ for various penetration depths $d^{13}$. For penetration
depths smaller than 35~mm the field distributions are only weakly localized and
$L_{loc}$ is not well defined.} \label{loclength}
\end{figure}

Despite the fact, that even for strongly localized modes the field distribution
is not symmetric with respect to $z_0$ -- mainly because of the small
dimensions of our system -- Eq.~(\ref{loclengtheq}) can be used to describe
{\it the envelope} of $\vec{E}^2$ for a wide range of penetration depths
$d^{13}$. In Fig.~\ref{loclength} one can see a nearly linear dependence
between $d^{13}$ and $L_{loc}$ for penetration depths above $35 {\rm mm}$. For
a smaller perturbation the field distribution shows a clear maximum (see e.g.
Fig.~\ref{single}b) around the perturbing screw, but Eq.~(\ref{loclengtheq})
cannot be used to describe the envelope of $\vec{E}^2$, since the decay is not
exponential, and $L_{loc}$ is therefore not well defined. To underline the
point that the field distribution has a maximum in the disturbed cell one
usually calls this form of localization {\it weak localization}
\cite{Kudrolli}. The behaviour of the field distributions of our microwave
resonator can be compared to the eigenfunctions computed for numerical models
with a single perturbation of a periodic chain \cite{Ishii}.

\begin{figure}[hbt]
\centerline{\epsfxsize=8.6cm \epsfbox{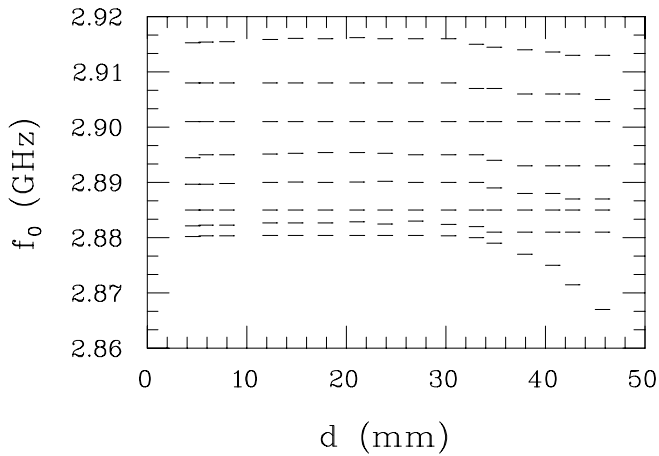}} \vspace*{2.5ex}
\caption{Resonance frequencies $f_0$ of the first eight modes for different
penetration depths $d^{13}$. A notable frequency shift which is going along
with a localization of the field distribution can only be observed for the
first mode.} \label{spaghetti}
\end{figure}

Beside the transition from an extended to an exponentially localized
eigenfunction the solutions of the Schr\"odinger equation do have other prominent
features in periodic potentials with a single perturbation. One of these is a
shift in the eigenvalue of the local impurity mode, while the eigenvalues of
the solutions in a band that are still extended over the system remain fixed.
An investigation of the resonance frequency $f_0$ of the first eight modes of
the microwave resonator while $d^{13}$ is increased, see Fig.~\ref{spaghetti},
shows that only the local impurity mode -- in this case the lowest excitation
of the cavity -- experiences a notable shift in its resonance frequency. Even
in the regime of weak localization, i.e. around $d^{13} \approx 30{\rm mm},$
where the field distributions of all modes are extended over the whole cavity,
no shift of the resonance frequency of the first mode is observed. The
transition from an extended or weakly localized state to a strong, i.e.
exponential, localized state can therefore not only be observed by measuring
the eigenfunctions but also by looking at the eigenfrequencies. To conclude
this section we like to state that we found a complete correspondence between
the eigenfunctions and eigenvalues of the Schr\"odinger equation in a periodic
potential with a single perturbation and the field distributions and resonance
frequencies of a periodic 3D microwave cavity with a single, sufficiently large
perturbation.

\begin{figure}[hbt]
\centerline{\epsfxsize=8.6cm \epsfbox{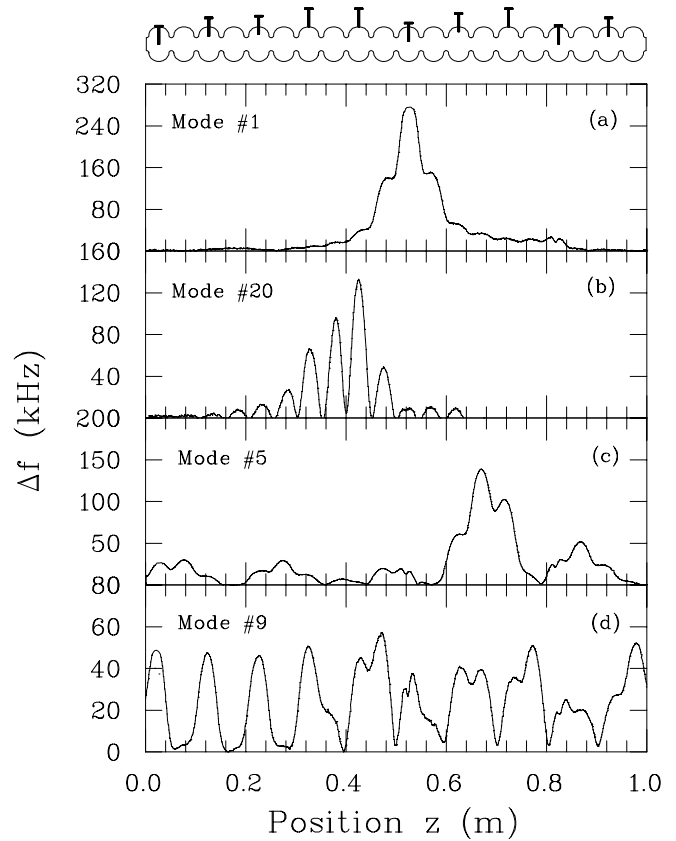}} \vspace*{2.5ex}
\caption{Field distributions of a disordered cavity -- the positions of the
screws are sketched above the plot. While on the edges of the first band the
modes are strongly localized ((a) and (b)) there are still weakly localized
modes (c) or even extended modes (d) in the middle of the band.}
\label{disturbed}
\end{figure}

\section{MULTIPLE PERTURBATIONS}
\label{multiple}

In contrast to the emerging of a local impurity mode at the position of a
single perturbation or impurity, one expects for a large class of
one-dimensional disordered systems that {\it all} the eigenfunctions of a
scalar wave equation are exponentially localized
\cite{Anderson,Pastur,Kunz,Delyon}. For disordered acoustic systems a
coexistence between extended, weakly localized and strongly localized modes has
been reported by He and Maynard \cite{Maynard}. To examine the behaviour of a
disordered electromagnetic system we set the penetration depths of all screws
to various, non periodic alternating lengths. The positions of the screws are
sketched in an inset above Fig.~\ref{disturbed} where also several different
field distributions are shown.

\begin{figure}
\centerline{\epsfxsize=8.6cm \epsfbox{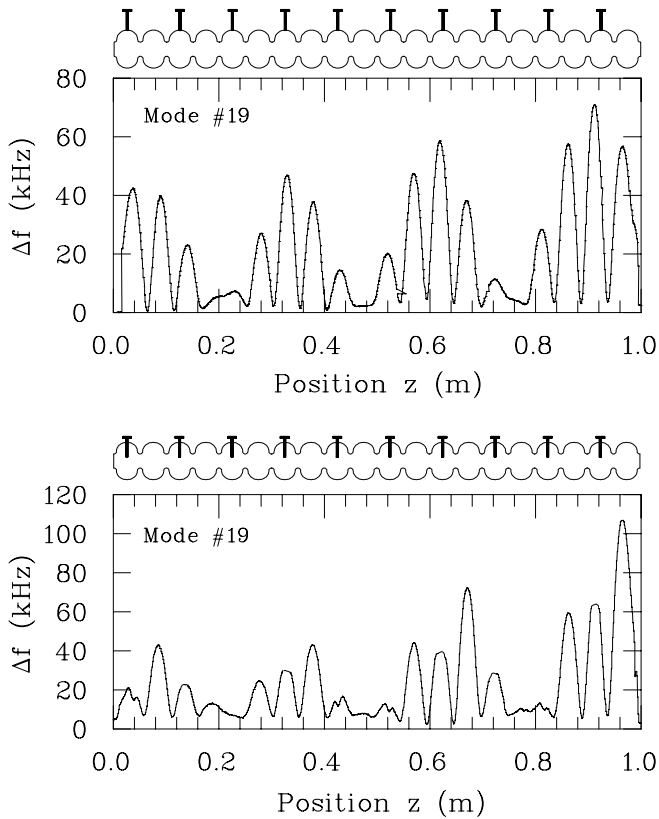}} \vspace*{2.5ex}
\caption{Field distribution of the 19th mode in the unperturbed cavity (top)
and in the {\it periodically} perturbed cavity (bottom). Both states are
extended over the whole resonator and have similar envelopes.}
\label{doubleperiod}
\end{figure}

In contrast to the predictions for one-dimensional systems in
\cite{Anderson,Pastur,Kunz,Delyon} not all field distributions are
exponentially localized as for example the first (Fig.~\ref{disturbed}a) or the
20th (Fig.~\ref{disturbed}b) mode with localization lengths of $L^1_{loc}
\approx 45{\rm mm}$ and $L^{20}_{loc} \approx 60{\rm mm}$. We still found
states which show weak localization, e.g. the fifth mode
(Fig.~\ref{disturbed}c), or that are still extended throughout the whole system
(ninth mode, Fig.~\ref{disturbed}d). Dean and Bacon found a similar behaviour
in an early numerical model where they studied the eigenfunctions of a
disordered harmonic chain composed of 22 light atoms and 28 heavy ones and
which is described by the Schr\"odinger equation \cite{Dean}. In their model,
localization occurs always on the upper edge of a band, while we are observing
localization at both, the upper and the lower edge of the first band. The
extended modes we are observing are found in the middle of the band like the
ninth mode shown in Fig.~\ref{disturbed}d. As mentioned above a coexistence of
localized and extended modes has been already observed for an acoustic system
\cite{Maynard}.

Our particular set-up allows the examination of another interesting system: If
the penetration depths off all screws are set to the same value, the system
will be periodic again -- an elementary cell is now composed of an unperturbed
and a perturbed cell -- and should therefore show only extended wave functions.
As an example we compare in Fig.~\ref{doubleperiod} the field distributions of
the 19th mode of the cavity with $d^i=0{\rm mm},\ i=1,3,\ldots,19$ and
$d^i=40{\rm mm},\ i=1,3,\ldots,19$. As expected we found, that both wave
functions have essentially the same envelope, despite the fact that for a {\it
randomly} perturbed setting the 19th mode is strongly localized. In both cases
the field distribution is not exactly symmetric -- an effect that is caused by
geometrical imperfections and which is also visible in the upper part of
Fig.~\ref{undisturbed}.

\section{CONCLUSION}

By using an appropriately shaped 3D microwave cavity we investigated a finite,
periodic or disordered system with eigenfunctions described by the vectorial
Helmholtz equation. The field distributions inside the cavity were measured by
analyzing the phaseshifts caused by a small dieelectric body inside the
resonator. In the case of a chain of quasi-identical cells we find extended
field distributions -- the whole resonator volume is excited.

It is well known that the eigenstates of a scalar wave equation are
exponentially localized in the case of infinitely long, disordered chains and
that there is always one local impurity mode per band in the case of a single
perturbation or impurity of the chain. Our experiments showed that the same
behaviour can be observed for a finite -- in fact with just 20 cells very small
-- system although it is described by the vectorial Helmholtz equation. We
observed the transition from an extended state to an exponentially localized
state in the case of a single perturbation by looking at the field distribution
and the resonance frequency of the mode. In the regime of strong localization
we found a linear dependence between the localization length and the
penetration depth of the screw that causes the perturbation. The current set-up
does not allow the investigation of totally disordered chains but enables us to
study systems where extended modes coexist with strong and weakly localized
ones which can be compared to certain numerical models for scalar problems
\cite{Dean} or results found in acoustic systems \cite{Maynard}. It is also
possible to study chains composed of a periodic array of two-cell elements in
which {\it all} states are extended over the whole cavity.

Our experiments showed that appropriately shaped microwave cavities exhibit all
the features of extended periodic systems -- despite the fact that they are
finite and in our case have to be described by vectorial wave equations. To our
knowledge this is also the first time that the localization of an eigenfunction
of the vectorial Helmholtz equation has been observed in an electromagnetic
system. The experimental verification of predictions on the statistical
behavior of extended billiard chains, see e.g. \cite{Doron}, should therefore
be possible and combine the results of the theory of periodic systems and
so-called quantum chaotic systems \cite{RichterBuch}. The experimental set-up
itself already allows the study of various problems from the fields of
solid-state physics and scattering problems in a very clean and pedagogical
way.

\section{acknowledgments}
We would like to thank the workshop of the Institut for Nuclear Physics in
Darmstadt for the excellent and precise modifications of the microwave
resonator. We would also like to thank T. Dittrich, S. Fishman, Y. Imry, C.
Rangacharyulu, U. Smilansky and H.-J. St\"ockmann for very helpful discussions.
This work has still been supported in part by the Sonderforschungsbereich 185
"Nichtlineare Dynamik" of the Deutsche Forschungsgemeinschaft (DFG) and through
the Forschergruppe with contract number DFG RI242/12-1.


\end{multicols}

\end{document}